\documentclass[osajnl,preprint,showpacs,superscriptaddress,12pt]{revtex4-1} 
\usepackage{amsmath,amssymb,graphicx}

\usepackage{dcolumn}
\usepackage{bm}
\usepackage{float}
\usepackage{enumitem}
\usepackage{url}

\newcommand\eq{\begin{equation}}
\newcommand\be{\begin{equation}}
\newcommand\eeq{\end{equation}}
\newcommand\ee{\end{equation}}
\newcommand\ar{\begin{eqnarray}}
\newcommand\ear{\end{eqnarray}}

\newcommand{\ii}{{\rm i}}

\renewcommand{\Re}{\mbox{Re}}

\begin{document}
\newcommand{\nm}{\mbox{ nm}}
\newcommand{\micron}{\mbox{ $\mu$m}}
\newcommand{\fm}{\mbox{ fm}}
\newcommand{\Hz}{\mbox{ Hz}}
\newcommand{\watt}{\mbox{ W}}
\newcommand{\m}{\mbox{ m}}
\newcommand{\cm}{\mbox{ cm}}
\newcommand{\km}{\mbox{ km}}
\newcommand{\seconds}{\mbox{ s}}
\newcommand{\minutes}{\mbox{ min}}
\newcommand{\volts}{\mbox{ V}}
\newcommand{\MeV}{\mbox{ MeV}}
\newcommand{\eV}{\mbox{ eV}}
\newcommand{\meV}{\mbox{ meV}}
\newcommand{\rad}{\mbox{ radians}}
\newcommand{\Kelvin}{\mbox{ K}}
\newcommand{\Siemen}{\mbox{ S}}
\newcommand{\angstrom}{\mbox{ $\AA$}}

\title{Broadband and efficient plasmonic control in the near-infrared and visible via strong interference of surface plasmon polaritons}

\author{C. H. Gan}\email{Corresponding author: c.h.gan@exeter.ac.uk}
\author{G. R. Nash}
\affiliation{College of Engineering, Mathematics and Physical Sciences, University of Exeter, Exeter EX4 4QF, United Kingdom}

\begin{abstract}Broadband and tunable control of surface plasmon polaritons in the near-infrared and visible spectrum is demonstrated theoretically and numerically with a pair of phased nanoslits. 
We establish with simulations supported by a coupled wave model that by dividing the incident power equally between two input channels, the maximum plasmon intensity deliverable to either side of the nanoslit pair is twice that for an isolated slit. 
For a broadband source, a compact device with nanoslit separation of the order of a tenth of the wavelength is shown to steer 
nearly all the generated plasmons to one side for the same phase delay, thereby achieving a broadband unidirectional plasmon launcher. The reported effect can be applied to the design of ultra-broadband and  efficient tunable plasmonic devices. 
\end{abstract}


\maketitle 

Plasmonics, the manipulation of light via sub-diffraction surface plasmon polariton (SPP) modes, has been exploited for various technologies over the last decade. Nanoscale plasmonic devices have been identified as potential platforms for the integration of high-speed photonic components and chip-scale electronic circuitry~\cite{zia_matl9}. To realize such plasmonic optoelectronic circuits, two of the key ingredients are broadband operation and active tunability. Tunable plasmonic devices based on phase-change 
materials~\cite{krasavin_apl84}, thermo-optic effects~\cite{bozhe_apl85,lereu_apl86}, electro-optic modulation~\cite{dicken_nl8,cai_nl9,qdeng_opex21}, ultrafast optical modulation~\cite{stockman_np3}, and selective mode excitation in a narrow slit~\cite{gan_opex12} have been proposed. 
Recently, polarization-based interference effects have been investigated for directional launching of SPPs in the visible~\cite{zayats_sci340,capasso_sci340}. 
In this study, we demonstrate theoretically and numerically broadband and tunable control of SPPs with a nanoslit pair by 
controlling the optical phase delay $\varphi$ in one of the slits. 
Our work is complementary to the investigations in Refs.~[10]~and~[11], and exploits the direct interference of SPPs scattered from the nanoslits by controlling the optical delay in one of the input arms. 
Additionally, for near-infrared frequencies where excitation of quasi-cylindrical waves (CWs) may dominate~\cite{lalanness}, we demonstrate that their contribution to the total electromagnetic fields is suppressed considerably in the nanoslit pair as they only interfere weakly in comparison to the SPPs. In this respect, the proposed scheme addresses spectral control to circumvent electromagnetic dispersion effects in the near-infrared, an issue that remains inadequately tackled in other similar doublet geometries proposed for directional launching of SPPs~\cite{qdeng_opex21,maier_opex20}.  
Another aspect uncovered in this study is that compared to the single slit device (such as in Refs.~[9]~and~[10]), the
SPP generation efficiency on either side of the nanoslit pair can be nearly doubled simply by dividing the incident power equally between two channels.

\begin{figure*}[ht!]
\vspace*{-0.2in} 
  \centering
\includegraphics[width=6.0 in]{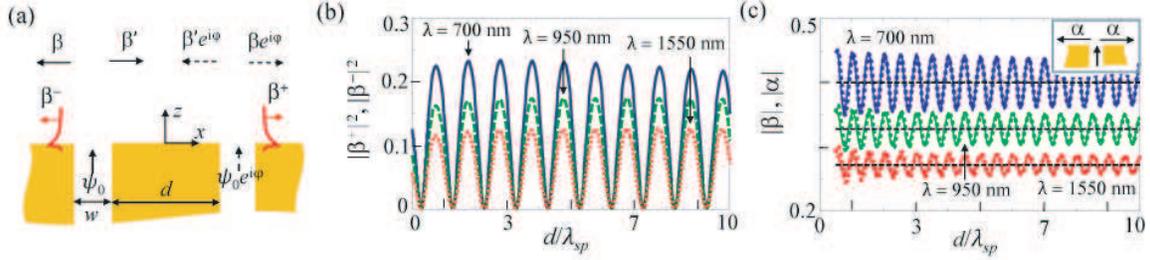} 
\vspace*{-0.4in} 
\caption{Geometry considered and definition of main physical parameters. (a) $\beta$ and $\beta'$ represent the scattered SPP amplitudes for the case either the left (black solid arrows) or right (black dashed arrows) slit is illuminated. $\beta^+$ and $\beta^-$ are the scattered SPP amplitudes directed to the right and left when both nanoslits are illuminated, with each carrying half the incident optical power. (b) SPP scattering efficiencies $|\beta^+|^2$ and $|\beta^-|^2$. (c) $\beta$ obtained from the coupled-wave model of Eq.~\eqref{model_cw} (solid curves) and simulations (dotted curves). Inset: definition of $\alpha$ (plotted as dashed horizontal lines). In (b) and (c), $\lambda = 700, 950,$ and $1550 \nm$, and $\varphi = 0$. 
}\label{Fig1}
\end{figure*}


The considered geometry consists of a pair of nanoslits of width $w$ in a metal plate separated by a distance $d$ as shown in Fig.~1(a), where each of the nanoslits is illuminated with the fundamental ${\rm TM_{0}}$ waveguide mode ($\psi_0$).  
The key parameters are the total scattered SPP amplitudes directed to the right or left of the nanoslits ($\beta^{\pm}$).
To describe the scattering of the SPP mode, we define 
for the case where {\em either one of the slits is illuminated}, the scattering amplitudes $\beta$ and $\beta'$ for the SPP propagating to infinity and that propagating towards the other slit, respectively.
Taking into the account the phase delay $\varphi$ in one of the input arms,
the amplitudes of the SPP mode launched to the right ($\beta^+$) and left ($\beta^-$) are
\eq\label{spbeta_exact}
\beta^+ = \frac{\beta e^{\ii \varphi} + \tau \beta' e^{\ii k_{sp}d'}}{\sqrt{2}}\,, \,\, \beta^- = \frac{\beta + \tau \beta' e^{\ii (k_{sp}d' + \varphi)}}{\sqrt{2}}\,,
\eeq
where $d' = d + w$, $\tau$ is the modal transmission coefficient of the SPP across a single slit, and the factor $\frac{1}{\sqrt{2}}$ takes into account that the amplitudes $\beta$ and $\beta'$ are normalized for unity Poynting flux of the mode $\psi_0$ in each of the slits. The SPP wavenumber $k_{sp} = \frac{2\pi}{\lambda}\sqrt{\frac{n_0^2 n_m^2}{n_0^2 +  n_m^2}} = k_r + \ii k_i$, where $\lambda$ is the illuminating wavelength, with $n_0$ and $n_m$ the refractive indices of the surrounding medium (taken to be free space, $n_0 = 1$) and the metal respectively. 
The SPP scattering amplitudes are calculated  with the mode orthogonality of translational-invariant lossy waveguides~\cite{lalanneprl}. For instance, $\beta^+ = \int_{-\infty}^{\infty}  \, [E_{z}(x, z)\,H^{{\rm SP-}}(x, z) - H(x, z)\,E_z^{{\rm SP-}}(x, z)] \,dz/\sqrt{2}\,,$ ($x > w+d/2$). Here $[H(x, z), E_z(x, z)]$ are the transverse field components of the total field scattered by the nanoslit pair, and $[H^{{\rm SP -}}(x, z)$, $E_z^{{\rm SP -}}(x, z)]$ are the analytically~\cite{Raether} calculated field components of the SPP mode propagating in the $-x$-direction with unit power-flow at $x = w+d/2$. 
All simulations of the electromagnetic fields are obtained with a fully-vectorial aperiodic-Fourier modal method~\cite{caojosa,besbes}, with the metal taken to be gold (data from Palik~\cite{palik}). The fields associated with the CWs on the metal surface are obtained by subtracting the SPP modal fields from the total field~\cite{lalanness}.
Unless otherwise specified, we will take $w \sim 0.23\lambda$ ($w = 160, 210,$ or $350 \nm$ for $\lambda = 700, 950,$ or $1550 \nm$) to optimize the SPP generation efficiency~\cite{lalanneprl}.
The separation $d$ is to be large enough to avoid mode splitting. 
For the fully symmetric case ($\varphi = 0$), Eq.~\eqref{spbeta_exact} actually remains valid even for the limiting case $d \rightarrow 0$, with the 
supermode of the nanoslit pair evolving gradually into the mode of a single slit of width $2w$. 
For $\varphi \neq 0$, Eq.~\eqref{spbeta_exact} is expected to remain reasonably accurate for separations as small as $d \sim 100 \nm$.

\begin{figure*}[ht!]
\vspace*{-0.2in}
  \centering
  \includegraphics[width = 6.0 in]{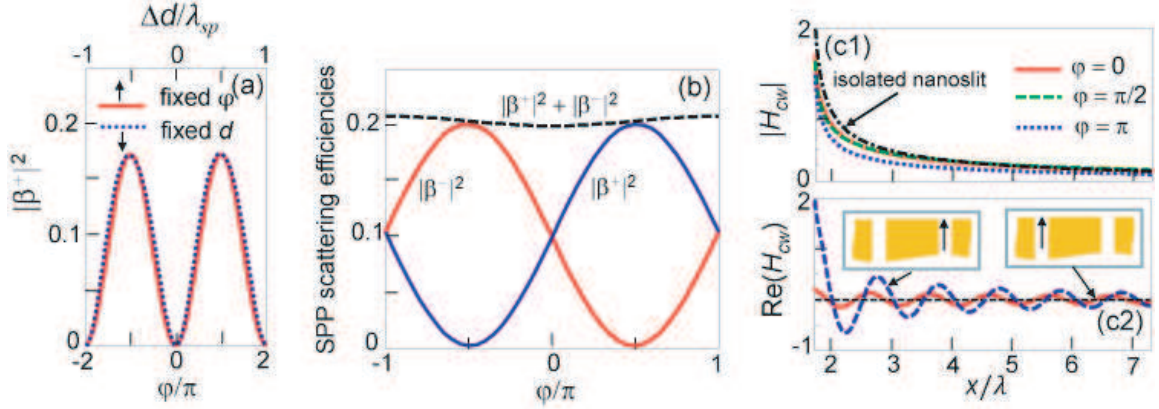}    
\vspace*{-0.4in}  
\caption{Dynamic control of SPPs and weak electromagnetic contributions from the CWs. (a) For a fixed separation ($d = 3.28\lambda_{sp}$), varying $\varphi$ is equivalent to laterally shifting the scattered SPP intensity. (b) Unidirectional launching of SPP at $\varphi = \pm \pi/2$. (c1) $|H_{cw}|$ for $\varphi = 0$, $\pi/2$, $\pi$, and for the case of the isolated slit. (c2) $\Re(H_{cw})$ launched to the right of the nanoslits when only either slit is illuminated (see insets). The horizontal line $\Re(H_{cw}) = 0$ is plotted for reference. In (b), (c1), and (c2), the slit separation $d$ is taken to be $3.03\lambda_{sp}$. In all cases, the wavelength $\lambda = 950 \nm$. }\label{Fig2}
  \end{figure*}

For nanoslit separation $d << L_{sp}$, where $L_{sp} = (2 k_{i})^{-1}$ is the SPP intensity decay length, the intensities of the SPPs launched to the right and left are approximately
\ar\label{beta_intensity}
|\beta^{\pm}|^2 & = & e^{-k_id} [\,|\beta|^2 + \tau^2 |\beta'|^2 \nonumber \\
&  & + 2\tau|\beta| |\beta'| \cos(k_r(d+w) \mp \varphi) \,] /2\,,
\ear
from which one sees that under the transformation $\varphi \longrightarrow -\varphi$, $|\beta^{+}|^2 = |\beta^{-}|^2$. Therefore any manipulation of the SPP intensity in one direction can be obtained for the SPP in the other direction by switching the sign of $\varphi$. 
In agreement with Eq.~\eqref{beta_intensity}, the simulated SPP intensities $|\beta^{\pm}|^2$ for three different 
wavelengths ($\lambda =  700 \nm$ (blue solid curve), $950 \nm$ (green dashed curve), and $1550 \nm$ (red dotted curve)) 
for $\varphi =0$ in Fig.~1(b) vary periodically with the 
effective SPP modal wavelength ($\lambda_{sp} = 2\pi/k_r$).
It can be seen that the SPPs interfere strongly among themselves, giving rise to near complete destructive interference at $d_{{\rm min}} = (2m+1)\lambda_{sp}/2 - w$ ($m$ being an integer).
Under constructive interference, the total SPP intensity  scattered in both directions reaches $46\%, 35\%$, and $25\%$ in order of ascending wavelength, which ranges from $45\%$ ($\lambda =  700 \nm$) to $75\%$ (near-infrared) more compared to an isolated slit. 
For small slit separations $d \lesssim 3\lambda_{sp}$, the peak SPP intensity is seen to decrease slightly. 
This slight decrease, together with the near complete destructive interference of the SPP, can be explained with the coupled wave model described below.

The model is based on fundamental SPP scattering coefficients for a single slit. 
In addition to $\tau$, we further define $\alpha$ as the SPP coupling coefficient of an isolated slit (inset of Fig.~1(c)), and $r$ the SPP modal reflection coefficient. 
According to the model,
\eq\label{model_cw}
\beta = \frac{\alpha (1 + ru(\tau v - r))}{(1-r^2 u)}, \quad \beta' = \frac{\alpha}{(1-r^2u)}\,,
\eeq
where $u = e^{\ii2k_{sp}d}$ and $v = e^{\ii k_{sp}w}$.
Calculations show that $|r|^2 < 0.1$ for the range of slit widths and wavelengths considered. The reflectivity gets increasingly weaker for increasing wavelengths where the longitudinal electric field component of the SPP becomes vanishingly small.
Due to the weak modal reflection, $\beta \approx \beta' \approx \alpha$, and $|\beta^{\pm}|^2 \approx |\alpha|^2[\,|1 + \cos(k_r (d+w) \mp \varphi)\,]$, ($\tau \approx 1$ for $w << \lambda$), i.e., {\em the maximum scattering efficiency of SPPs can be nearly doubled via interference by equally dividing the incident power 
between two identical channels.} 
Figure~1(c) depicts the agreement between the model of Eq.~\eqref{model_cw} and the simulated results. 
Towards the near-infrared, the SPP amplitudes oscillate more noticeably for small separations, departing from the pure SPP coupled-wave model. This is characteristic of contributions 
from the cross-conversion~\cite{lalanne_xcon} of CWs that are more strongly excited at longer wavelengths. 
Similar trends are found for $\beta'$ (not shown), which is nearly identical to $\alpha$. 
As the term $ru$ in Eq.~\eqref{model_cw} becomes more significant for small separations $d$, $|\beta|$ and $|\beta'|$ generally deviate more from each other for decreasing $d$, explaining the slight decrease in the peak SPP intensities for small $d$ (see Fig.~1(b)). Additionally, as $r$ is small, the phase difference between $\beta$ and $\beta'$, taken to be zero in Eq.~\eqref{beta_intensity}, can be neglected.

\begin{figure*}[ht!]
\vspace*{-0.2in} 
  \centering
  \includegraphics[width = 6.0 in]{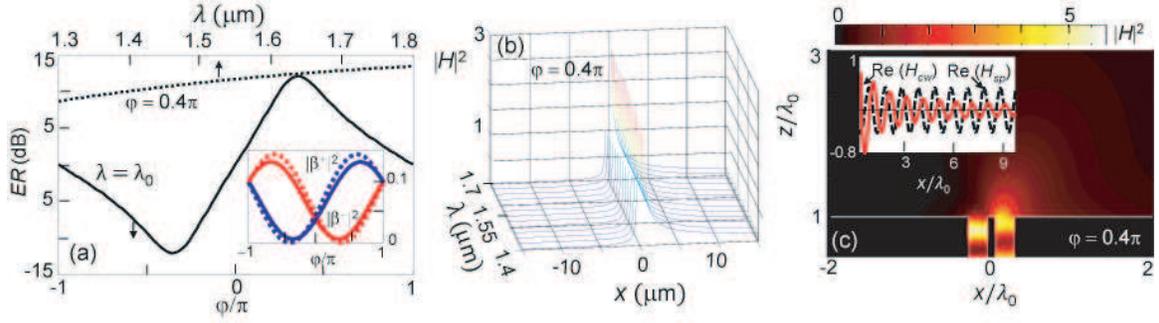}  
\vspace*{-0.4in}    
\caption{Near-dispersionless unidirectional launching of SPPs with a compact device ($d = 150 \nm$) for central operating wavelength $\lambda_0 = 1550 \nm$. (a) $ER$ for varying $\varphi$ for $\lambda_0$, and for $\varphi = 0.4 \pi$ for $1300 \nm \leq \lambda \leq 1800 \nm$. Inset: $|\beta^+|^2$ (blue) and $|\beta^-|^2$ (red) for $\lambda_0$ obtained from simulations (solid curves) and the coupled-wave model (dotted curves). (b) Intensity of the total magnetic field $|H^2|$ launched to either side of the nanoslits on the metal surface 
for $1400 \nm \leq \lambda \leq 1700 \nm$. (c) Spatial distribution of $|H^2|$ for $\lambda_0$. Inset: $\Re(H_{sp})$ and $\Re(H_{cw})$ launched to the right on the metal surface. The horizontal line $\Re(H_{sp}) = \Re(H_{cw}) = 0$ is plotted for reference.}\label{Fig3}
\end{figure*}

For a nanoslit pair with a fixed separation, Eq.~\eqref{beta_intensity} indicates that the scattered SPP intensity can be shifted 
by an amount $\Delta d/\lambda_{sp} = \varphi/2\pi$, as demonstrated 
in Fig.~2(a) for $\lambda = 950 \nm$, $w  = 210 \nm$, and $d = d_{{\rm min}}|_{m = 3} = 3.28\lambda_{sp}$. 
For $\varphi = 0$, the generated SPPs are trapped as standing waves between the nanoslits. 
On switching $\varphi$ to $\pi$, $35\%$ of the incident power is launched as SPPs in both 
directions ($|\beta^{\pm}|^2 = 0.175$). 
By making the choice $d = (2m+1)\lambda_{sp}/4 - w$ instead, unidirectional launching of SPPs can be achieved by switching $\varphi$ from $-\pi/2$ to $\pi/2$, as
illustrated in Fig.~2(b) for the case $d = 3.03\lambda_{sp}$. 
The maximum efficiency of SPPs on either side is $20\%$, twice that of the isolated slit ($|\alpha|^2 = 10\%$), in excellent agreement with the coupled wave model. Due to near-complete destructive interference of the SPPs, a near-unity modulation 
depth $(|\beta^{\pm}|^2_{\rm max}-|\beta^{\pm}|^2_{\rm min})/|\beta^{\pm}|^2_{\rm max}$ ($\sim 99\%$ in Fig.~2(b)) is obtainable for virtually any pair of phased-nanoslits. 
Let us also note that by applying the phase delay to both slits, one may tune the nanoslit pair to a desirable initial state (see Fig.~2(a)), and then employ the phase delay in the other slit for full dynamic control of the SPPs.

Figure~2(c1) shows the amplitude $|H_{cw}|$ of the magnetic component of the CWs  for $\varphi = 0$ (solid red curves), $\varphi = \pi/2$ (dashed green curves), 
and $\varphi = \pi$ (dotted blue curves). In contrast to the SPPs, 
the CWs decay quickly to an approximately uniform low level after propagating $\sim 2 - 3\lambda$ from the nanoslits independently of $\varphi$.
This is mainly due to the lack of spatial coherence~\cite{mandel} 
between the part of the CW that is directly scattered from the ${\rm TM_{0}}$ mode, and of that transmitted across the other slit. 
The latter is much weaker in terms of magnitude and have a generally arbitrary phase relationship with respect to the directly excited CWs~\cite{lalanness,ganprb83}, as evident in Fig.~2(c2) showing $\Re(H_{cw})$ launched to the right of the nanoslits when only either slit is illuminated. 
Due to symmetry, these are indeed the CWs scattered to the left and right by a singly-illuminated pair, but are shown superposed along the same spatial coordinates for comparison.

Next, we demonstrate a compact ($d \sim \lambda/10$) unidirectional SPP launcher operating over a spectral bandwidth of $\sim 300 \nm$ when illuminated by a uniform {\em white} source with a central wavelength $\lambda_0 = 1550 \nm$. 
Let us take the nanoslit separation $d$ to be $150 \nm$, and define the extinction ratio $ER = 10\log(|\beta^+|^2/|\beta^-|^2)$. 
Figure~3(a) shows $ER$ for $\lambda_0$ with varying $\varphi$, and for $\varphi = 0.4 \pi$ for $1300 \nm \leq \lambda \leq 1800 \nm$. 
For $\lambda_0$ and $\varphi = 0.4 \pi$, $ER$ reaches a maximum of 12~dB with an efficiency $|\beta^+|^2 = 11\%$ (see inset Fig.~3(a)). The predicted efficiency is about 1.5 times instead of twice that of the isolated slit ($|\alpha|^2 \sim 7\%$) mainly because the maxima and minima of $|\beta^+|^2$ and $|\beta^-|^2$ do not coincide. 
From Eq.~\eqref{beta_intensity}, the maximum and minimum of $|\beta^+|^2\,(|\beta^-|^2)$ lie near $\varphi = +(-)0.65\pi$ and $\varphi = -(+)0.35\pi$, respectively. 
For sufficiently short separations $d$, the near-constant phase factor $k_r(d+w)$ suggests resilience of the nanoslit pair to electromagnetic dispersion. Figure~3(a) shows that, for $\varphi = 0.4 \pi$, $ER$ remains greater than 10~dB for  $1400 \nm \leq \lambda \leq 1700 \nm$. 
The intensity of the total magnetic field at the metal surface is shown in Fig.~3(b), showing that the nanoslit pair can steer all the SPPs generated by the broadband source to the right (or the left) with a fixed value of $\varphi = 0.4 \pi\,(-0.4 \pi)$ across the $300 \nm$ spectral range. 
Figure~3(c) shows that apart from a radiation lobe directed towards the right, all the energy is either directed to the SPPs or back to the nanoslits. 
It is worth noting that for $x \geq 2\lambda_0$, the launched field consist mainly of the SPP mode rather than the CW (see inset Fig.~3(c))
even though the latter is expected to be dominant for $x \lesssim \lambda |n_m|^2/(2\pi) \approx 15\lambda$~\cite{lalanness}. 
While the separation $d$ is taken to be $150 \nm$ in this instance, it can be as small as necessary for precluding effects of mode-splitting between the nanoslits. With this being the lower limit, near-dispersionless unidirectional launching can be achieved for arbitrary separations $d << \lambda_{sp}$ with the present scheme. 
This offers additional flexibility in terms of device design over doublet geometries that are restricted to a discretized separation distance for directional launching~\cite{qgong_2013} or such as that illustrated in Fig.~2(b) contrived for a single wavelength.

In summary, we have demonstrated broadband and tunable plasmonic control with a phased nanoslit pair. Remarkably, by dividing the power equally into two channels, the nanoslit pair can deliver up to twice the plasmons in either direction in comparison to an isolated slit. 
Due to the strong interference of the SPPs, spurious contributions from the CWs are heavily suppressed even in the near-infrared. The proposed geometry, which highly resembles a 50:50 beamsplitter, is ideal for integration with next-generation optoelectronic circuitries~\cite{zia_matl9,krenn_apl}. Alternatively, the optical delay $\varphi$ which is in the order of the effective SPP wavelength could, for example be incorporated using integrated tunable optical filters~\cite{lenz_jqe37}. 
The predicted broadband directional launching of SPPs could additionally open up new possibilities for plasmonic-based signal processing at telecom wavelengths.

CHG acknowledges stimulating comments from P. Lalanne. This work was supported by the Engineering and Physical Sciences Research Council (grant number EP/J011932/1).

\end{document}